\documentclass[aps,orb,twocolumn,amssymb,showpacs]{revtex4}

\usepackage{psfig}
\usepackage{graphicx}

\begin{document}

\def\be{\begin{equation}}
\def\ee#1{\label{#1}\end{equation}}

\title{Palatini approach to $1/R$ gravity and its implications to the 
late Universe }
\author{Gilberto M. Kremer}\email{kremer@fisica.ufpr.br} 
\author{Daniele S. M. Alves}\email{dsma01@fisica.ufpr.br} 
\affiliation{Departamento de F\'\i sica, Universidade Federal do Paran\'a,
Caixa Postal 19044, 81531-990 Curitiba, Brazil}

\begin{abstract}
By applying the Palatini approach to the $1/R$-gravity model it is possible
to explain the present accelerated expansion of the Universe. Investigation
of the late Universe limiting case shows that: 
(i)  due to the curvature effects the energy-momentum tensor of the 
matter field is not covariantly conserved; (ii) however, it is possible to
reinterpret the curvature corrections as sources of the gravitational field,
by defining a modified energy-momentum tensor; (iii) with the adoption of this
modified energy-momentum tensor the Einstein's field equations are recovered
with two main modifications: the first one is the weakening of 
the gravitational effects of matter whereas the second is the emergence of an
effective varying "cosmological constant"; 
(iv) there is a transition in the evolution of the cosmic scale factor from
a power-law scaling $a\propto t^{11/18}$ to an asymptotically exponential 
scaling $a\propto \exp(t)$; (v) the energy density 
of the matter field scales as $\rho_m\propto (1/a)^{36/11}$; (vi)
the present age of the Universe and the  
decelerated-accelerated  transition redshift
are smaller   than the corresponding ones in the $\Lambda$CDM model.
\end{abstract}
\maketitle

\section{Introduction}

The interpretation of recent cosmological observations concerning the 
redshifts of supernovae of type Ia (see e.g.~\cite{Per}) suggests
that the Universe has evolved from a past decelerated phase to a present 
accelerated period. The decelerated period is normally described by a 
matter dominated Universe, and in order to simulate the transition from 
a decelerated to an accelerated phase from Einstein's field equations one 
has to introduce another cosmological fluid which is gravitationally
self-repulsive.  This fluid with  negative pressure  is identified 
with some form of dark energy which dominates over the matter field 
in the accelerated period. At least two candidates for dark energy show up
in literature, namely, the quintessence and the Chaplygin gas
(see e.g.~\cite{ST}, \cite{KMP} and the references therein, respectively).
Another way to obtain the decelerated-accelerated transition without the need
of dark energy is to introduce  curvature corrections into
Einstein's field equations (see e.g.~\cite{CCT,Vol, Meng} and the 
references therein).

The present work is based on the method proposed by Vollick~\cite{Vol}
which makes use of the Palatini approach in order to derive
the field equations from a modified Hilbert-Palatini action. 
In the  gravitational part of the 
action figures the sum of two contributions: one of them is the usual  term 
proportional to the curvature scalar whereas the other -- which represents 
a correction -- is inversely proportional to it. The other part is the action
for the matter.

Among other results -- for the late times in the Universe when the curvature 
corrections become relevant  -- we show that 
the energy-momentum tensor of the matter field is not covariantly 
conserved due to curvature effects,  but it is possible to define a
modified energy-momentum tensor of the sources which is covariantly conserved.
The definition of the modified energy-momentum tensor is a reinterpretation
of the curvature corrections as sources of the gravitational field, and it 
turns out to be the sum of two constituents: a matter field whose coupling to 
the space-time geometry is weakened and an effective varying "cosmological
constant". By considering curvature effects up to the 
second order approximation  we also show that 
the cosmic scale factor evolves from a matter dominated period with
$a\propto t^{11/18}$ to a cosmological constant dominated phase where
$a\propto \exp(t)$, while  the energy density 
of the matter field scales as $\rho_m\propto (1/a)^{36/11}$.
Moreover, we compare the results of the present
work with those that follow from the $\Lambda$CDM model. In particular, it is
shown that due to curvature effects the present age of the Universe and the 
redshift in the decelerated-accelerated transition
are smaller  than the corresponding ones in the $\Lambda$CDM model.

\section{Field equations}

We follow the work of Vollick~\cite{Vol} and write the modified
Hilbert-Palatini action as
\be
S=\int\left[-{{\cal L}(R)\over 2\kappa}+{\cal L}_M\right]\sqrt{-g}d^4x.
\ee{1}
Here, $\kappa=8\pi G$, ${\cal L}_M$ is the Lagrangian density of the 
matter field and ${\cal L}(R)$ refers to a generic Lagrangian density that 
depends on the scalar curvature $R=g^{\mu\nu}{R^\sigma}_{\mu\sigma\nu}$, 
where ${R^\sigma}_{\mu\lambda\nu}$ is the Riemann tensor given by:
\be
{R^\sigma}_{\mu\lambda\nu}=\partial_{\nu}\Gamma^\sigma_{\mu\lambda}
-\partial_{\lambda}\Gamma^\sigma_{\mu\nu}
+\Gamma^\alpha_{\mu\lambda}\Gamma^\sigma_{\nu\alpha}
-\Gamma^\alpha_{\mu\nu}\Gamma^\sigma_{\lambda\alpha}.
\ee{2b}

The Palatini approach yields the field equations
\be
{\cal L}^\prime(R)R_{\mu\nu}-{1\over2}{\cal L}(R)g_{\mu\nu}=-
\kappa T_{\mu\nu},
\ee{2a}
by performing the variation of the action with respect to $ g_{\mu\nu}$.
Above $R_{\mu\nu}={R^\sigma}_{\mu\sigma\nu}$ is the Ricci tensor,
the prime in the term ${\cal L}^\prime(R)$ refers to a differentiation 
with respect to $R$ and the energy-momentum  tensor $T_{\mu\nu}$ of the
matter field is given by
\be
T_{\mu\nu}=-{2\over\sqrt{-g}}{\delta({\cal L}_M\sqrt{-g})\over\delta 
g^{\mu\nu}}.
\ee{2}
Furthermore, the variation of the action with respect to 
$ \Gamma_{\mu\nu}^\sigma$ leads to an equation which reduces to 
\be
\nabla_\sigma({\cal L}^\prime(R)g^{\mu\nu}\sqrt{-g})=0.
\ee{3}

It follows from equation (\ref{3}) that the affine connection is the 
Christoffel symbol with respect to the metric $h_{\mu\nu}\equiv{\cal L}^\prime
(R)g_{\mu\nu}$:
\be
\Gamma^\sigma_{\mu\nu}={1\over2}h^{\alpha\sigma}\left[\partial_\mu 
h_{\nu\alpha}+\partial_\nu h_{\alpha\mu}-\partial_\alpha h_{\mu\nu}\right].
\ee{3a}
The relationship between the 
affine connection $ \Gamma_{\mu\nu}^\sigma$ and the Riemannian connection of
the  metric $g_{\mu\nu}$, denoted by $\tilde \Gamma_{\mu\nu}^\sigma$, 
is given by
\be
\Gamma_{\mu\nu}^\sigma=\tilde\Gamma_{\mu\nu}^\sigma+{1\over2{\cal L}^\prime}
\left[2\delta^\sigma_{(\mu}\partial_{\nu)}{\cal L}^\prime-g^{\sigma\tau}
g_{\mu\nu}\partial_\tau{\cal L}^\prime\right].
\ee{4}
From equation  (\ref{4}) it follows a relationship between 
the Ricci tensor $R_{\mu\nu}$ and the  Riemannian Ricci tensor 
$\tilde R_{\mu\nu}$, constructed from the Riemannian connection
$\tilde \Gamma_{\mu\nu}^\sigma$, that reads
$$
R_{\mu\nu}=\tilde R_{\mu\nu}-{3\over 2({\cal L}^\prime)^2}
\partial_\mu{\cal L}^\prime\partial_\nu{\cal L}^\prime
$$
\be
+{1\over {\cal L}^\prime}\tilde\nabla_\mu\tilde\nabla_\nu{\cal L}^\prime
+{1\over 2{\cal L}^\prime}\tilde\nabla^\sigma\tilde\nabla_\sigma
{\cal L}^\prime g_{\mu\nu}.
\ee{5}
Above, $\tilde \nabla_\mu$ denotes the covariant differentiation associated 
with the Riemannian connection $\tilde \Gamma_{\mu\nu}^\sigma$.

The field equations (\ref{2a}) can be rewritten in terms of the Riemannian
Ricci tensor $\tilde R_{\mu\nu}$ and of the Riemannian scalar curvature 
$\tilde R$ of the $g_{\mu\nu}$ metric, 
thanks to (\ref{5}), as
\be
\tilde R_{\mu\nu}-{1\over2}\tilde Rg_{\mu\nu}=-\kappa \tilde T_{\mu\nu},
\ee{7}
where the modified energy-momentum tensor of the sources $\tilde T_{\mu\nu}$ 
becomes
$$
\tilde T_{\mu\nu}={T_{\mu\nu}\over{\cal L}^\prime}-{1\over\kappa}\left[
{3\over 2({\cal L}^\prime)^2}\partial_\mu{\cal L}^\prime
\partial_\nu{\cal L}^\prime-{1\over {\cal L}^\prime}
\tilde\nabla_\mu\tilde\nabla_\nu{\cal L}^\prime
\right.
$$
\be
\left.
-{1\over 2{\cal L}^\prime}\tilde\nabla^\sigma\tilde\nabla_\sigma
{\cal L}^\prime g_{\mu\nu}+{1\over2}\left({{\cal L}\over{\cal L}^\prime}-
\tilde R\right)g_{\mu\nu}
\right].
\ee{8}
From the Bianchi identity one can get the conservation law 
for the modified energy-momentum tensor of the sources, i.e., 
$\tilde\nabla_\nu \tilde T^{\mu\nu}=0$. 

We stress that, although the conformal metric $h_{\mu\nu}$ figures in the 
expression (\ref{3a}) for the affine connection $\Gamma^\sigma_{\mu\nu}$, 
we do not regard it, but $g_{\mu\nu}$, as the physical metric, since it was 
with respect to $g_{\mu\nu}$ that we performed the variation of the action 
in order to obtain the field equations, and it is $g_{\mu\nu}$ that figures 
in the definition of the energy-momentum tensor of the matter fields, eq. 
(\ref{2}).

From now on we shall restrict ourselves to 
a homogeneous and isotropic plane Universe
described by the Robertson-Walker metric $ds^2=dt^2-a(t)^2(dx^2+dy^2+dz^2)$
where $a(t)$ represents the cosmic scale factor. 

We write the energy-momentum tensor of the matter field as
${T^\mu}_\nu=\hbox{diag}(\rho_m,-p_m,-p_m,-p_m)$ where $\rho_m$ and $p_m$
denote its energy density and pressure, respectively.
Moreover, we represent the modified energy-momentum tensor of the sources as
${{\tilde T}^\mu}_{\;\;\,\nu}=\hbox{diag}(\rho,-p,-p,-p)$. 
From the expression (\ref{8}) one can obtain that the energy density of 
the  modified energy-momentum tensor of the sources reads
$$
\rho={\rho_m\over {\cal L}^\prime}-{1\over \kappa}\left[{3\over 
2({\cal L}^\prime)^2}(\partial_0{\cal L}^\prime)^2-{3\over 
2({\cal L}^\prime)}\partial^2_0{\cal L}^\prime\right.
$$
\be
\left.
-{3\over 
2({\cal L}^\prime)}{\dot a\over a}\partial_0{\cal L}^\prime+
{1\over2}\left({{\cal L}\over{\cal L}^\prime}-\tilde R\right)\right],
\ee{9}
whereas its pressure is given by
$$
p={p_m\over {\cal L}^\prime}-{1\over \kappa}\left[{1\over 
2({\cal L}^\prime)}\partial^2_0{\cal L}^\prime
\right.
$$
\be
\left.
+{5\over 
2({\cal L}^\prime)}{\dot a\over a}\partial_0{\cal L}^\prime
-{1\over2}\left({{\cal L}\over{\cal L}^\prime}-\tilde R\right)\right].
\ee{10}

The acceleration and Friedmann equations follow from the field equations
(\ref{7}), yielding
\be
{\ddot a\over a}=-{\kappa\over6}(\rho+3p),
\qquad
\left({\dot a\over a}\right)^2={\kappa\over3}\rho.
\ee{11}
The evolution equation for the energy density can be obtained 
from the two above
equations or from the conservation law 
for the modified energy-momentum tensor of the sources
$\tilde\nabla_\nu \tilde T^{\mu\nu}=0$ in a comoving frame, and reads
\be
\dot \rho+3{\dot a\over a}(\rho+p)=0.
\ee{12}
The expressions (\ref{11}) and (\ref{12}) are the usual acceleration, 
Friedmann and energy equations for a flat Universe whose source is a 
fluid with energy density $\rho$ and pressure $p$.

\section{Curvature corrections}

Let us explore the modified Lagrangian density which has a term proportional
to the scalar curvature plus a term which is inversely proportional to it. 
We follow Vollick~\cite{Vol} and write
\be
{\cal L}=R-{\alpha^2\over3R},
\ee{13}
where $\alpha$ is a constant.

From the trace of the field equations (\ref{2a}) together with (\ref{13}) 
one can obtain the  following expression for the scalar curvature in terms 
of the trace of the energy-momentum tensor of the matter field 
$T=g^{\mu\nu}T_{\mu\nu}$:
\be
R=\alpha\left[{\kappa T\over2\alpha}\pm\sqrt{1+\left({\kappa T\over2\alpha}
\right)^2}\right].
\ee{14}

The Einstein field equations are recovered in the limit $2\alpha/(\kappa T)
\ll1$ and by choosing the plus sign if $T>0$ or the minus sign if $T<0$. In 
that case $R\approx\kappa T$, implying that $2\alpha/R\ll1$ and 
${\cal L}^\prime\approx1$, and therefore, from (\ref{7}) and (\ref{8}) the 
Einstein equations directly follow. It is in this approximation that the 
Newtonian limit is verified, since in any measurement we could perform to 
check it we would have $\kappa T\gg2\alpha$. The opposite limit, 
$\kappa T\ll2\alpha$, can only be observed in cosmological scales in which 
$T$ is of order $10^{-26}$kg/m$^3$.

Here we are interested in matter fields such that $p_m\leq\rho_m/3$. Assuming 
that the weak energy condition~\cite{HE} holds ($\rho_m\geq0$), we will always 
have that $T=\rho_m-3p_m\geq0$ and therefore the plus sign in equation 
(\ref{14}) will be chosen.

From now on we shall restrict ourselves to the limiting case  where
${\kappa T/(2\alpha)}\ll1$ so that  up to 
the second order approximation we have
\be
R\simeq\alpha\left[1+{\kappa T\over2\alpha} \right],
\ee{15}
and the Lagrangian density and its derivative with respect to $R$ become
\be
{\cal L}\simeq{2\over 3}\alpha\left[1+{\kappa T\over\alpha}
\right],\qquad
{\cal L}^\prime\simeq{4\over 3}\left[1-{\kappa T\over4\alpha}
 \right],
\ee{16}
respectively. 

By considering the above approximations one can write 
the modified field equations (\ref{7}) as
\be
\tilde R_{\mu\nu}-{1\over2}\tilde Rg_{\mu\nu}=-{3\over 4}\kappa  
T_{\mu\nu}-\Lambda g_{\mu\nu}, 
\ee{7a}
thanks to (\ref{8}), (\ref{15}) and (\ref{16}). In the above equation 
we have introduced a cosmological constant term
\be
\Lambda={3\alpha\over8}-{R\over8}={\alpha\over4}-{\kappa T\over16}.
\ee{7b}

The right-hand side of equation (\ref{7a}) has a simple physical 
interpretation:
(i) in the first term the coupling between the matter with the space-time 
geometry is decreased by a factor 3/4 due to curvature effects; 
(ii) the second term is  an effective cosmological constant $\Lambda$
which is a sum of a legitimate cosmological constant $\alpha/4$ 
plus a varying "cosmological constant" $-\kappa T/16$ which 
decreases in module with time.
We stress that the two terms of the cosmological 
constant are due to the curvature effects as it can be seen from the  
first equality in equation (\ref{7b}).

The Bianchi identity $\tilde\nabla_\nu(\tilde R_{\mu\nu}-\tilde 
Rg_{\mu\nu}/2)=0$ leads to
\be
\tilde\nabla_\nu{T^\nu}_\mu=-{4\over 3\kappa}\partial_\mu \Lambda,
\ee{7c}
showing that the energy-momentum tensor of the matter field is not 
a covariantly conservative quantity. The right-hand side of (\ref{7c}) 
may be rewritten as $-{4\partial_\mu \Lambda/ (3\kappa)}={\partial_\mu T/12}=
{\partial_\mu R/ (6\kappa)}$ thanks to (\ref{7b}), hence the non-conservation
of the energy-momentum tensor of the matter field is a consequence of curvature
effects.

The evolution equation for the energy density of the matter field in 
a comoving frame follows from equation (\ref{7c}), yielding
\be
\dot \rho_m+3{\dot a\over a}(\rho_m+p_m)={1\over 12}\dot T
={1\over6\kappa}\dot R.
\ee{18}
One can infer from the above equation that the curvature drains energy 
from the matter field or equivalently the matter 
field transfer more energy to the gravitational field due to curvature effects.

The energy density and pressure of the  modified energy-momentum 
tensor of the sources $\tilde T_{\mu\nu}$ are given by
\be
\rho={3\over 4}\rho_m+\left({\alpha\over 4\kappa}-{1\over16}T\right)
\equiv{3\over 4}\rho_m+\rho_\Lambda,
\ee{17a}
\be
p={3\over 4}p_m-\left({\alpha\over 4\kappa}-{1\over16}T\right)
\equiv{3\over4}p_m+p_\Lambda,
\ee{17}
thanks to (\ref{9}),  (\ref{10}), (\ref{15}) and (\ref{16}). 
Hence, the energy density of the  
modified energy-momentum tensor of the sources  can be interpreted as
a sum of two terms. The first one is a linear function of the energy density 
of the matter field where the factor 3/4 is related to the weakening of 
the gravitational influence of the matter field due to the curvature effects.
The second term is the energy density associated with the effective 
cosmological constant  $\rho_\Lambda=-p_\Lambda$. We note that evolution
equation for the matter field (\ref{18}) is recovered if we insert the
expressions (\ref{17a}) and (\ref{17}) into the evolution equation (\ref{12})
for energy density of the modified energy-momentum tensor of the sources.

From now on we shall consider a Universe filled with dust
so that  $p_m=0$ and hence the trace of the energy-momentum
tensor of the matter reduces to $T=\rho_m$. In this case from the evolution
equation (\ref{18}) for the energy density of the matter field it follows that
\be
\rho_m=\rho_m^0\left({a_0\over a}\right)^{3/\beta},
\ee{19}
where $\beta=11/12$. The parameter $\beta$ was introduced in order to compare
the case where the curvature effects are considered ($\beta=11/12$) with
the case where the curvature effects are not taken into account ($\beta=1$). 

We infer from (\ref{19}) that the energy density of the matter field 
scales as $(a_0/a)^{36/11}$ and therefore
decays more rapidly in comparison with $(a_0/a)^3$ when $a_0/a>1$. 
It is worth to say 
that for a radiation field the trace of the energy momentum tensor vanishes, 
and there is no effect of the curvature on the radiation field, 
since  from equation (\ref{18}) it follows
that its energy density scales as $(a_0/a)^4$.

We introduce the critical density and the density parameters usually defined by
\be
\rho_c={3H^2\over\kappa},
\qquad\Omega_m={\rho_m\over\rho_c},\qquad
\Omega_\Lambda={\rho_\Lambda\over\rho_c},
\ee{21}
where $H=\dot a/a$ denotes the Hubble parameter. Hence one obtains from
(\ref{17a}) that $\Omega_{\rm eff}+\Omega_\Lambda=1$,
where $\Omega_{\rm eff}=3\Omega_m/4$ is the effective density parameter of the 
matter field which takes into account the curvature effects. Moreover, 
one can introduce a density parameter for the legitimate cosmological 
constant $\tilde\Omega_\Lambda=\alpha/(4\kappa\rho_c)$ so that it follows 
the relationship 
$\tilde\Omega_\Lambda=1-\beta\Omega_{\rm eff}$. We note that for the
$\Lambda$CDM model $\beta=1$ and  we have 
to consider $\Omega_{\rm eff}\equiv\Omega_m$.

The Friedmann equation (\ref{11})$_2$ can now be rewritten thanks to 
(\ref{19}) 
as
\be
\left({\dot a\over a}\right)^2=\tilde\Omega^0_\Lambda H_0^2\left[1
+{\beta\Omega_{\rm eff}^0\over\tilde\Omega_\Lambda^0}\left({a_0\over a}
\right)^{3/\beta}\right],
\ee{22}
and the general solution of this differential equation reads
\be
{a(t)\over a_0}=\left[\sqrt{\beta\Omega_{\rm eff}^0\over\tilde\Omega_\Lambda^0}
\sinh\left({3\over2\beta}\sqrt{\tilde\Omega_\Lambda^0}H_0 t\right)
\right]^{2\beta/3}.
\ee{23}

\begin{figure}\begin{center}
\includegraphics[width=8.5cm]{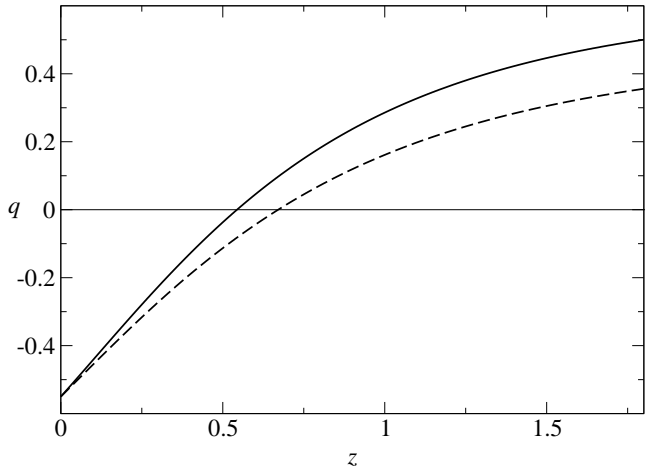}
\caption{Deceleration parameter $q$ vs redshift $z$. With curvature 
effects - straight line; $\Lambda$CDM model - dashed line. }
\end{center}\end{figure}

In the limiting case where  $t\ll1/(\sqrt{\tilde\Omega_\Lambda^0}H_0)$ we 
get from equation (\ref{23}) a matter dominated Universe which evolves 
according to
\be
{a(t)\over a_0}\simeq \left({3\over2}\sqrt{\beta\Omega_{\rm eff}^0}H_0 t
\right)^{2\beta/3}.
\ee{24}
Hence,  for a matter dominated Universe where the curvature effects 
are taken into account  $a\propto t^{11/18}$  and it expands more slowly  
than a matter dominated Universe without curvature 
effects since in the last case  $a\propto t^{2/3}$.

In the limiting case where $t\gg1/(\sqrt{\tilde\Omega_\Lambda^0}H_0)$ 
we obtain from (\ref{23}) a cosmological constant dominated Universe 
with an accelerated expansion which evolves according to
\be
{a(t)\over a_0}\simeq \left({\beta\Omega_{\rm eff}^0\over 4\tilde
\Omega_\Lambda^0}\right)^{\beta/3}\exp\left(\sqrt{\tilde\Omega_\Lambda^0}
H_0 t\right).
\ee{25}
 
From equation (\ref{23}) one can obtain the relationship 
\be
t_0H_0={2\beta\over3}{1\over \sqrt{1-\beta\Omega_{\rm eff}^0}}\hbox{arsinh}
\left(\sqrt{1-\beta\Omega_{\rm eff}^0\over\beta\Omega_{\rm eff}^0}\right),
\ee{26}
where $t_0$ denotes the present time. 
If we consider that  
$\Omega_{\rm eff}^0= 0.3$ and $\beta=11/12$ it follows from (\ref{26})
that $t_0H_0\approx 0.905$. In the $\Lambda$CDM model we would have  
$\beta=1$ and get $t_0H_0\approx 0.964$. Hence due to curvature effects 
the age of the Universe becomes smaller  than the one where the 
curvature effects are not taken into account.

The deceleration parameter $q$ has the same expression as
the one without considering the curvature effects, i.e.,
\be
q\equiv -{\ddot a\over aH^2}={3\over2}\Omega_{\rm eff}-1,
\ee{27a}
whereas the density parameter as a function of the redshift $z$ reads
\be
\Omega_{\rm eff}={\Omega_{\rm eff}^0(1+z)^{3/\beta}\over1+
\beta\Omega_{\rm eff}^0(1+z)^{3/\beta}-\beta\Omega_{\rm eff}^0}.
\ee{27}

\begin{figure}\begin{center}
\includegraphics[width=8.5cm]{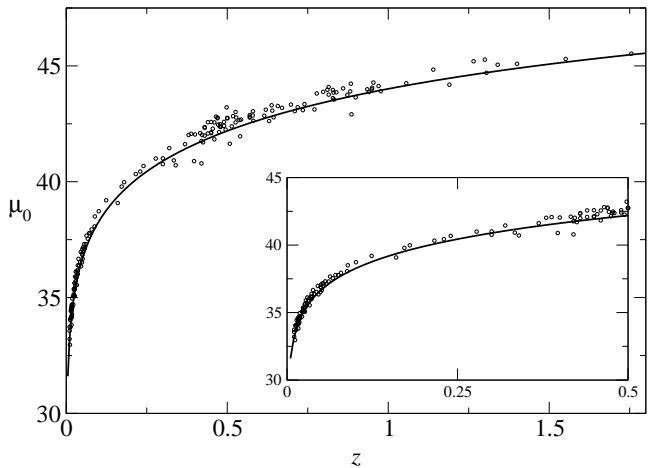}
\caption{Difference $\mu_0$ vs redshift $z$.}
\end{center}\end{figure}

We have  combined  equations (\ref{27a}) and (\ref{27}) and plotted in 
figure 1 the deceleration parameter $q$ as a function of the redshift $z$ for 
$\Omega_{\rm eff}^0= 0.3$. The case where the curvature effects 
are taken into account ($\beta=11/12$) is represented by a straight line
while the $\Lambda$CDM model ($\beta=1$) is 
represented by a dashed line. From these curves we infer that the deceleration
parameter increases more rapidly with the increasing of the redshift 
for case where the curvature effects are present, 
since $\Omega_{\rm eff}$ also shows this
behavior due to the relationship $\rho_m\propto(1+z)^{3/\beta}$.
Moreover, one obtains that the transition from a 
decelerated epoch to an accelerated one takes place at a redshift 
$z\approx 0.544$ when the curvature effects are taken into account 
and at $z\approx 0.671$ in the $\Lambda$CDM model.

From the luminosity distance, which is given by
\be
d_L=(1+z)H_0^{-1}\int_0^z{dz\over\sqrt{1+\beta\Omega_{\rm eff}^0
(1+z)^{3/\beta}-\beta\Omega_{\rm eff}^0}},
\ee{28}
one can build the difference between the apparent magnitude $m$ and the 
absolute magnitude $M$ of a source: 
$\mu_0=m-M=5\log d_L+25$ where 
$d_L$ is given in Mpc.

In figure 2 we have plotted the difference $\mu_0$
as function of the redshift $z$ -- by considering $\Omega_{\rm eff}^0= 0.3$
and $cH_0^{-1}=3000/0.72$ Mpc --
for the case where the curvature effects
are taken into account. The circles in this figure represent 
the experimental values taken from the work by Riess et al.~\cite{Ri} 
for 185 data of supernovae 
of type Ia. We note that the curve fits the values of the data 
at an acceptable level.
The curve which corresponds to the $\Lambda$CDM model is not sketched 
in this figure, since its departure from the curve plotted is not too  
significant concerning the dispersion of the values of the data.

As a final remark we note that additional modifications emerge when
higher order approximations in $\kappa T/(2\alpha)\ll1$ are considered, such as
the dependence of the equations on higher order derivatives of the trace $T$
and the time-variation of $\beta$.



\begin{thebibliography}{99}


\bibitem{Per}   S. Perlmutter et al.  { Astrophys. J.} {\bf 517}, 565 (1999);
A. G. Riess et al.  {Astrophys. J.} {\bf 560}, 49 (2001);
 M. S.  Turner and   A. G. Riess, {Astrophys. J.} {\bf 569}, 18 (2002).

\bibitem{ST} R. R. Caldwell, R. Dave and P. J. Steinhardt   
{ Phys. Rev. Lett} {\bf 80}, 1582 (1998);
I. Zlatev, L. Wang and P. J. Steinhardt   
{ Phys. Rev. Lett} {\bf 82}, 896 (1999);
G. M. Kremer and  F. P. Devecchi  { Phys. Rev. D} 
{\bf  67}, 047301 (2003); 
P. J. E. Peebles  and  B. Ratra,   {Rev. Mod. Phys.}
{\bf 75}, 559 (2003).


\bibitem{KMP} A. Yu. Kamenshchik, U. Moschella and V. Pasquier, V.
{ Phys. Lett. B} {\bf 511}, 265 (2001);
J. C. Fabris, S. V. B.  Gon\c calves and  P. E. de Souza,  
 { Gen. Relativ. Gravit.} {\bf 34}, 53 (2002);
M. C. Bento, O. Bertolami and A. A. Sen   { Phys. 
Rev. D}  {\bf 66}, 043507 (2002);
A. Dev, J. S.  Alcaniz and  D. Jain  { Phys. Rev. D}  
{\bf 67}, 023515 (2003);  G. M. Kremer,    Gen. Relat. Grav.  
{\bf 35}, 1459 (2003).


\bibitem{CCT}  S. Capozziello, S. Carloni,  and  A. Trois,  astro-ph/0303041. 

\bibitem{Vol} D. N. Vollick, Phys. Rev. D {\bf 68}, 063510 (2003).

\bibitem{Meng} X. H. Meng and P. Wang, Class. Quantum Grav. {\bf 21}, 951 
(2004).
 
\bibitem{HE} S. W. Hawking and G. F. R. Ellis, {\it The Large Scale Structure 
of Spacetime} (Cambridge University Press, Cambridge, England, 1973).

\bibitem{Ri} A. G. Riess et al. astro-ph 0402512.






\end{thebibliography}
\end{document}